\documentclass[floatfix,showpacs,preprint,preprintnumbers,amsmath,amssymb,prb]{revtex4}
\usepackage{graphicx}
\usepackage{dcolumn}
\usepackage{bm}

\begin{document}

\preprint{}

\title{Penetration depth, multiband superconductivity, and absence of muon-induced perturbation in superconducting PrOs$_{\bm{4}}$Sb$_{\bm{12}}$}

\author{Lei Shu\footnote{Current address: Department of Physics and Institute for Pure and Applied Physical Sciences, University of California, San Diego, La Jolla, California 92093}}
\author{D.~E. MacLaughlin}
\author{W.~P. Beyermann}
\affiliation{Department of Physics, University of California, Riverside, California 92521}
\author{R.~H. Heffner}
\affiliation{Los Alamos National Laboratory, Los Alamos, New Mexico 87545}
\author{G. D. Morris}
\affiliation{TRIUMF, 4004 Wesbrook Mall, Vancouver, B.C., Canada
V6T 2A3}
\author{O. O. Bernal\footnote{Current address: Division of Materials Research, National Science Foundation, Arlington, VA 22230}}
\affiliation{Department of Physics and Astronomy, California State University, Los Angeles, California 90032}
\author{F. D. Callaghan}
\author{J. E. Sonier}
\affiliation{Department of Physics, Simon Fraser University, Burnaby, B.C., Canada V5A 1S6}
\author{W.~M. Yuhasz}
\author{N.~A. Frederick}
\author{M.~B. Maple}
\affiliation{Department of Physics and Institute for Pure and Applied Physical Sciences, University of California, San Diego, La Jolla, California 92093}
\date{\today}

\begin{abstract}
Transverse-field muon spin rotation (TF-$\mu$SR) experiments in the heavy-fermion superconductor PrOs$_{4}$Sb$_{12}$ ($T_{c}=1.85$ K) suggest that the superconducting penetration depth $\lambda(T)$ is temperature-independent at low temperatures, consistent with a gapped quasiparticle excitation spectrum. In contrast, radiofrequency (rf) inductive measurements yield a stronger temperature dependence of $\lambda(T)$, indicative of point nodes in the gap. Muon Knight shift measurements in the normal state of PrOs$_{4}$Sb$_{12}$ suggest that the perturbing effect of the muon charge on the neighboring Pr$^{3+}$ crystalline electric field is negligibly small, and therefore is unlikely to cause the difference between the TF-$\mu$SR and rf results. The discrepancy appears to be related to multiband superconductivity in PrOs$_{4}$Sb$_{12}$.
\end{abstract}


\maketitle

\section{\label{sect:Intro}Introduction}

In the phenomenon of multiband superconductivity (MBSC), which was first treated theoretically in 1959 using BCS theory,\cite{SMW59} distinct energy gaps open up on different sheets of the Fermi surface in the superconducting state. One of the well-studied multiband superconductors is the binary intermetallic compound MgB$_{2}$,\cite{GDUS02,CEDJ03, TYTK03} which has two bands and two superconducting gaps. Based on a quasiclassical theory, microscopic calculations of the electronic structure in the vortex state have been made in a two-band superconducting model.\cite{NIM02, IMNM04} These suggest that at low applied fields the dominant contribution to the change of the total density of states (DOS) comes from the small-gap band, which gives rise to spatially extended quasiparticles (QPs). At high fields these loosely bound states become delocalized, with the vortex core size determined by the more localized states associated with the larger gap. The presence of delocalized QPs modifies the DOS, transport properties associated with the quasiparticles, and the spatial field distribution around a vortex.\cite{GySc91}

Recently strong evidence for MBSC has been found in the filled skutterudite~PrOs$_{4}$Sb$_{12}$ from thermal transport measurements in the vortex state.\cite{SBMF05, SBMB06} PrOs$_{4}$Sb$_{12}$ has attracted much attention since its superconductivity was discovered in 2002.\cite{BFHZ02} It is the only known Pr-based heavy-fermion superconductor ($T_{c}=1.85$~K) and exhibits a number of extraordinary properties.\cite{MHZY03} The Pr$^{3+}$ ground state is a nonmagnetic $\Gamma_{1}$ singlet, which is separated from a $\Gamma_{4}^{(2)}$ first excited state (tetrahedral notation\cite{THY01}) by only $\sim$10~K\@.\cite{MHZF02,ANOS02,KINM03,GOBM04} A novel ordered phase appears at high fields and low temperatures. There is evidence that time-reversal symmetry is broken in the superconducting state.\cite{ATKS03} Radiofrequency (rf) inductive measurements of the magnetic penetration depth~$\lambda$ in the Meissner state,\cite{CSSS03} thermal conductivity measurements in a rotated magnetic field,\cite{INGM03} and flux-line lattice distortion\cite{HMID04} all suggest nodes in the superconducting gap. A double superconducting transition has been observed in specific heat measurements.\cite{MHZF02,VFPvL03,MBFS04a,GDPvL06} However, recent specific heat and heat transport measurements\cite{SBMB06} on a highly homogeneous single crystal show only one transition peak in the specific heat and a fully-gapped Fermi surface. The latter result is corroborated by angle-dependent specific heat measurements on single crystals.\cite{SYCY07} Previous transverse-field muon spin rotation (TF-$\mu$SR) measurements of $\lambda$ in the vortex state of PrOs$_{4}$Sb$_{12}$ found evidence for a BCS-like activated dependence at low temperature,\cite{MSHB02} suggesting the absence of gap nodes. Thus there are a number of open questions concerning the superconducting order parameter in PrOs$_{4}$Sb$_{12}$.

The present article reports a detailed TF-$\mu$SR study of PrOs$_{4}$Sb$_{12}$. TF-$\mu$SR experiments\cite{Brew94} have proved invaluable in characterizing both the superconducting and normal states of superconducting materials;\cite{SBK00,Soni07} in particular, a TF-$\mu$SR study\cite{CLKS05} of superconducting NbSe$_2$ clearly revealed effects of MBSC on the vortex-state field distribution in this compound. We compare TF-$\mu$SR and rf inductive penetration depth measurements in superconducting PrOs$_{4}$Sb$_{12}$, and discuss the previously-reported\cite{SMHC06} discrepancy between these measurements in terms of extreme MBSC in PrOs$_{4}$Sb$_{12}$. Preliminary reports of parts of this work have been published.\cite{SMHC06,MSHS08} 

We have also carried out TF-$\mu$SR measurements of the Knight shift in the normal state of PrOs$_{4}$Sb$_{12}$, which suggest that the perturbing effect of the positive-muon ($\mu^{+}$) charge on the neighboring Pr$^{3+}$ crystalline electric fields (CEF) is negligibly small. This indicates that the muon charge is unlikely to be the source of the discrepancy between the TF-$\mu$SR and rf results. A model calculation for the perturbed CEF energy levels, based on the approach of Kaplan, Schenck, and co-workers,\cite{FAGG95a, FAGG95b,FAGS94,TAGG97} is described. Higemoto {\em et al.}\cite{HSKO07} have reported a TF-$\mu$SR study of PrOs$_{4}$Sb$_{12}$, with emphasis on the Knight shift in the superconducting state. 

The remainder of this introduction contains three brief pedagogical sections: a description of the elements of the TF-$\mu$SR technique used in this study (Sec.~\ref{sect:TFmusr}), a review of the important features of the $\mu^{+}$ Knight shift (Sec.~\ref{sect:Knight-intro}), and an introduction to the CEF model calculation (Sec.~\ref{sect:CEF}). After a description of the experimental procedure (Sec.~\ref{sect:exp}), in Sec.~\ref{sect:results} we describe our experimental results in PrOs$_{4}$Sb$_{12}$, which include the temperature dependence of the TF-$\mu$SR relaxation rate and penetration depth, magnetic susceptibility data, and normal-state $\mu^{+}$ Knight shift measurements. The implications of these results for the nature of both the superconducting and the normal state are discussed in Sec.~\ref{sect:discussion}\@. We summarize our results in Sec.~\ref{sect:conc}.

\subsection{\label{sect:TFmusr}Transverse-field muon spin rotation in the vortex state}

In the TF-$\mu$SR technique\cite{Brew94} spin-polarized positive muons ($\mu^{+}$) are implanted in the sample and precess in a local magnetic field applied perpendicular to the initial $\mu^{+}$ polarization. This precession is detected using the asymmetry of the $\mu^{+}$ beta decay (the decay positron is emitted preferentially in the direction of the $\mu^{+}$ spin). The distribution of $\mu^{+}$ precession frequencies directly reflects the distribution of magnetic fields in the sample. Thus TF-$\mu$SR can be used to measure the magnetic field distribution of the vortex lattice in a type-II superconductor and the local magnetic susceptibility in the normal state.

In the vortex lattice each vortex possesses a normal-state-like core with radius of order of the superconducting coherence length $\xi$, surrounded by a shielding supercurrent. These supercurrents give rise to an inhomogeneous magnetic field that is periodic in the vortex lattice. An analytical Ginzburg-Landau (GL) model for the spatial field profile~$\mathbf{B}(\mathbf{r})$ (Ref.~\onlinecite{YDdRB97}) yields 
\begin{equation}
 \label{eq:distribution}
 \mathbf{B}(\mathbf{r})=B_{0}(1-b^{4})\sum_{K}\frac{e^{-i\mathbf{K}\cdot \mathbf{r}}uK_{1}(u)}{\lambda^{2}K^{2}} \hat{\mathbf{z}},
\end{equation}
where the $\mathbf{K}$ are the reciprocal lattice vectors of the vortex unit cell, $K_{1}(u)$ is a modified Bessel function, $b=B/B_{c2}$ is the reduced field, and
\begin{equation}
 \label{eq:u}
 u^{2}=2\xi^{2}K^{2}(1+b^{4})[1-2b(1-b)^{2}],
\end{equation}
with $\xi$ the GL coherence length. This result is valid in the limit~$\lambda^2 K_{\mathrm{min}}^2 \gg 1$, where $K_{\mathrm{min}}$ is the smallest nonzero reciprocal lattice vector.

TF-$\mu$SR is a sensitive probe of this field distribution.\cite{SBK00} The functional form of the $\mu^{+}$ spin relaxation, which depends on the field distribution, is fit to the functional form~$G(t)\cos(\omega_\mu t+\phi)$, where the frequency $\omega_\mu$ and phase $\phi$ describe the average $\mu^{+}$ precession and the relaxation function $G(t)$ describes the loss of phase coherence due to the distribution of precession frequencies. The relaxation rate associated with $G(t)$ is a measure of the rms width $\delta B$ of this distribution. The expression\cite{Bran88}
\begin{equation}
 \label{eq:distri}
 \delta B^{2}(T)=0.00371\Phi_{0}^{2}\lambda^{-4}(T),
\end{equation}
where $\Phi_{0}$ is the flux quantum, then gives a rough estimate of $\lambda$ for a triangular vortex lattice in the London limit ($\lambda \gg \xi$). More accurate relations are available,\cite{SBK00,Bran03,Soni07} but in the present case lead to negligible corrections to Eq.~(\ref{eq:distri}). 

To estimate the field distribution width without fitting a theoretical model, a Gaussian distribution of local fields is often assumed. Then the time dependence of the $\mu^{+}$ spin polarization is proportional to $\exp(-{\textstyle\frac{1}{2}}\sigma^{2}t^{2})$, where $\sigma = \gamma_{\mu}\delta B$ is the $\mu^{+}$ relaxation rate; $\gamma_{\mu}$ is the $\mu^{+}$ gyromagnetic ratio. It has been pointed out\cite{SBK00} that this procedure is approximate at best, because the field distribution is not expected to be Gaussian. In a more microscopic approach, a lineshape analysis program has been written to fit the GL model to TF-$\mu$SR data.\cite{SBK00,Soni07} The GL model is preferred over the London model because it self-consistently accounts for the vortex cores. In this fitting program the fit function is the Fourier transform of the magnetic-field distribution function~$n(B') = \int d^2r\,\delta[B' - B(\mathbf{r})]$, which can be obtained from Eq.~(\ref{eq:distribution}). These fits utilize the entire form of the field distribution, not just its second moment. They yield an effective `$\mu$SR' penetration depth $\lambda_{\mu\mathrm{SR}}$, which becomes the true penetration depth only after extrapolation to $B = 0$.\cite{Soni07} 

Results from such fits in PrOs$_{4}$Sb$_{12}$ (Ref.~\onlinecite{SMHC06}) are described below in Sec.~\ref{sect:TF}.

\subsection{\label{sect:Knight-intro} Muon Knight shift}

In TF-$\mu$SR the total field at the $\mu^{+}$ site is given by the sum of the applied field~$H$, the internal field induced by the applied field, and the demagnetization and Lorentz fields.\cite{Amat97} The relative $\mu^{+}$ frequency shift
\begin{equation}\label{eq:knight shift 1}
K_{\mu}^{\ast}=\frac{\omega_{\mu}}{\omega_{\mathrm{ref}}} - 1,
\end{equation}
where $\omega_{\mathrm{ref}}= \gamma_{\mu}H$ is the $\mu^{+}$ frequency in ``free space" (i.e., no condensed matter effects), must be corrected for the contribution of the demagnetization and Lorentz fields $K_{\mathrm{DL}}$ to obtain the $\mu^{+}$ Knight shift\cite{Sche85}
\begin{equation}\label{eq:knight-DL}
K_{\mu}=K_{\mu}^{\ast}-K_{\mathrm{DL}},
\end{equation}
which contains the relevant information about the local magnetic susceptibility.

In a paramagnetic metal, $K_{\mu}$ originates from hyperfine fields produced by the field-induced polarization of conduction electrons and localized electronic moments. The contribution from the conduction electrons is temperature independent and is usually very small, of the order of 100 ppm.\cite{Sche85} The local moments, in the present work the Pr$^{3+}$ ions, contribute to $K_{\mu}$ via two coupling mechanisms: (1)~the dipolar interaction between the local moments and the $\mu^{+}$, which may be described as a dipolar field at the $\mu^{+}$ interstitial site, and (2)~the indirect RKKY interaction,\cite{Sche85} in which an additional spin polarization of the conduction electrons due to the local moments produces a hyperfine contact field at the interstitial $\mu^{+}$. Both contributions are proportional to the local-moment susceptibility:
\begin{equation}\label{eq:knight shift 2}
K_{\mu}^{i}=(A_{\mathrm{dip}}^{ii}+A_{\mathrm{con}})\chi_{i} \quad (i = x,y,z)
\end{equation}
in the principal-axis coordinate system of the dipolar tensor~$\bm{\mathsf{A}}_{\mathrm{dip}}$, where the $A_{\mathrm{dip}}^{ii}$ are the diagonal elements of $\bm{\mathsf{A}}_{\mathrm{dip}}$ and $A_{\mathrm{con}}$ is the contact hyperfine coupling constant, assumed isotropic. For a cubic lattice ($\chi_x = \chi_y = \chi_z$), only the hyperfine contact field at the interstitial $\mu^{+}$ site contributes to the average shift~$K_{\mu}=\frac{1}{3}\sum_{i}K_{\mu}^{i}$, since the contribution from the dipole-dipole interaction vanishes $\left(\sum_i A_{\mathrm{dip}}^{ii} = 0\right)$.

The local susceptibility is sensitive to change in the Pr$^{3+}$ CEF, and the $\mu^{+}$ charge may induce such a change. If so, the modified local susceptibility will be reflected in a breakdown of the proportionality of the $\mu^{+}$ Knight shift to the measured bulk susceptibility $\chi^{\mathrm{bulk}}$ since in this case $\chi_{i}\neq \chi^{\mathrm{bulk}}$. This effect has been studied in detail by Kaplan, Schenck, and co-workers\cite{FAGG95a,FAGG95b,FAGS94,TAGG97} in Pr-based compounds with singlet Pr$^{3+}$ CEF ground states. As noted above and discussed in Sec.~\ref{sect:CEF}, our results in PrOs$_{4}$Sb$_{12}$ show little if any such effect.

\subsection{\label{sect:CEF} CEF model calculation}

The crystal structure of PrOs$_{4}$Sb$_{12}$ belongs to the $Im\bar{3}$ space group, with Pr$^{3+}$ ions at the points of a bcc unit cell. As an initial approximation we assume the CEF Hamiltonian ${\mathcal{H}}_{\mathrm{CEF}}$ for the Pr$^{3+}$ ions has $O_{h}$ (cubic) point group symmetry. Then
\begin{equation}\label{eq:hami1}
{\mathcal{H}}_{\rm
CEF}=B_{4}^{0}O_{4}^{0}+5B_{4}^{0}O_{4}^{4}+B_{6}^{0}O_{6}^{0}-21B_{6}^{0}O_{6}^{4},
\end{equation}
where $O_{4}^{0}$, $O_{4}^{4}$, $O_{6}^{0}$, and $O_{6}^{4}$ are the Stevens operators for a given angular momentum quantum number~$J$, and the $B$s are parameters usually determined from experiment.\cite{LLW62} ${\mathcal{H}}_{\mathrm{CEF}}$ splits the Pr$^{3+}$ $J = 4$ Hund's-rule multiplet into a $\Gamma_1$ singlet, a $\Gamma_3$ doublet, and two triplets ($\Gamma_{4}$ and $\Gamma_{5}$). Pr$^{3+}$ ions in PrOs$_{4}$Sb$_{12}$ actually have $T_{h}$ (tetrahedral) point group symmetry, however,\cite{THY01} for which an additional sixth-order term appears in the CEF Hamiltonian [Eq.~(\ref{eq:hami1})]. This term mixes the $\Gamma_{4}$ and $\Gamma_{5}$ triplet wave functions with each other\cite{THY01} but has a relatively small effect on the CEF energies.\cite{GOBM04} If it is predominant, a $\Gamma_{4}$ or $\Gamma_{5}$ triplet will be the ground state, inconsistent with the singlet ground state in nonmagnetic PrOs$_{4}$Sb$_{12}$. However, if this term is small, the change in the physical properties in zero magnetic field can be approximated by changing the parameters in the cubic CEF Hamiltonian [Eq.~(\ref{eq:hami1})] slightly. Thus we use the cubic CEF Hamiltonian for simplicity.

In the presence of an external magnetic field, the Zeeman interaction mixes and splits the CEF energy levels. The magnetic susceptibility is given by\cite{TAGG97}
\begin{equation}\label{eq:sus}
\chi_{\rm CEF}=\frac{\sum_{n}\left[(E_{n}^{(1)})^{2}/kT-2E_{n}^{(2)}\right]\exp\left(-E_{n}^{(0)}/kT\right)}{\sum_{n}\exp\left(-E_{n}^{(0)}/kT\right)},
\end{equation}
where the $E_{n}^{(0)}$ are the unperturbed cubic CEF levels, $E_{n}^{(1)}=\mu_{B}g\langle\phi_{n}|J|\phi_{n}\rangle$ with $\phi$ the CEF wave functions and $g$ the Land{\'e} g-factor, and 
\begin{equation}
E_{n}^{(2)}=\sum_{n^{'}\neq
n}\mu_{B}^{2}g^{2}\frac{|\langle\phi_{n}|J|\phi_{n}^{'}\rangle|^{2}}{E_{n}^{(0)}-E_{n^{'}}^{(0)}}.
\end{equation}
In the molecular-field approximation, the measured magnetic susceptibility $\chi$ is given by $\chi=\chi_{\mathrm{CEF}}/(1 - \ell\chi_{\mathrm{CEF}})$, where $\ell$ is the molecular field parameter that describes exchange interactions between Pr$^{3+}$ ions.\cite{MuMa57}

Equation~(\ref{eq:sus}) shows that the observed susceptibility is directly related to the local CEF energy levels. As noted above, if we take into account possible muon-induced change of the CEF, we must consider a modified CEF Hamiltonian to calculate the resultant change of the Pr$^{3+}$ local susceptibility.\cite{FAGG95a,FAGG95b,FAGS94,TAGG97} In PrOs$_{4}$Sb$_{12}$ the Pr$^{3+}$ first excited state is separated from the singlet ground state by only $\sim$10~K\@.\cite{MHZF02,ANOS02,KINM03,GOBM04} It is particularly important to determine possible muon-induced changes in the local CEF energy levels, since CEF excitations are central to a number of theories of superconductivity in PrOs$_{4}$Sb$_{12}$.\cite{MKH03,KMS06} Changes in CEF splitting might therefore affect local superconducting properties such as the vortex-state supercurrent.

\section{Experiments} \label{sect:exp}

TF-$\mu$SR experiments were carried out at the M15 and M20 channels at TRIUMF, Vancouver, Canada, on a mosaic of oriented PrOs$_{4}$Sb$_{12}$ crystals. The crystals were mounted on a thin GaAs backing, which rapidly depolarizes muons in transverse field and minimizes any spurious signal from muons that do not stop in the sample. TF-$\mu$SR asymmetry data\cite{Brew94} were taken for temperatures in the range 0.02--250~K and $\mu_{0}H$ between 10~mT and 1.0~T applied parallel to the $\langle$100$\rangle$ axes of the crystals. For the $\mu^{+}$ Knight shift measurements, the applied field was determined by measuring the precession frequency of muons that stopped in a small piece of silver foil included with the sample.

For the superconducting-state TF-$\mu$SR measurements the magnetic field~$H$ was applied in the normal state, and the sample was cooled in constant field to below $T_c$. For PrOs$_{4}$Sb$_{12}$ $H_{c1}(0) \lesssim 45$~Oe,~\cite{HFZB03,CMSF05} so that $H$ was significantly larger than $H_{c1}$ and the sample was always in the vortex-lattice state. The demagnetizing field~$-4\pi DM$, where $D$ is the demagnetization coefficient and $M < 0$ in the superconducting state,  adds to the field~$B = H + 4\pi(1-D)M$ inside the sample, rendering it closer to the applied field. In Sect.~\ref{sect:knight} we estimate an effective $D \approx 0.8$, which yields an estimated variation of $B$ with $H_{c1}$ (i.e., with temperature) of only a few percent. The strong bulk vortex pinning in PrOs$_4$Sb$_{12}$ (Ref.~\onlinecite{CMSF05}) also suppresses changes in $B$. Thus modification of the vortex-state field distribution by flux expulsion, demagnetization effects, etc., seems unlikely to play an appreciable role in the experiments.

\section{Experimental Results} \label{sect:results}

\subsection{\label{sect:TF}Superconducting-state TF-$\mathbf{\mu}$SR and magnetic penetration depth}

In this section we describe the temperature and field dependencies of the TF-$\mu$SR relaxation data in superconducting PrOs$_4$Sb$_{12}$, and compare the superconducting penetration depth~$\lambda(T)$ obtained from these data with inductive measurements~\cite{CSSS03} in the Meissner state. We concentrate on $\lambda(T)$ at low temperatures, where power-law behavior is evidence for gap nodes and temperature-independent (or activated) behavior signals a fully-gapped Fermi surface. The behavior of the data at higher temperatures is more complicated and will not be considered 
in detail. Furthermore, as noted above in Sec.~\ref{sect:TFmusr}, the effective TF-$\mu$SR penetration depth is expected to be field dependent, and approximates the true value only as $H \rightarrow 0$.\cite{Soni07} Thus we concentrate on the results at low temperatures and (for the TF-$\mu$SR data) low fields.

\begin{figure}[ht]
 \begin{center}
 \includegraphics[width=0.45\textwidth]{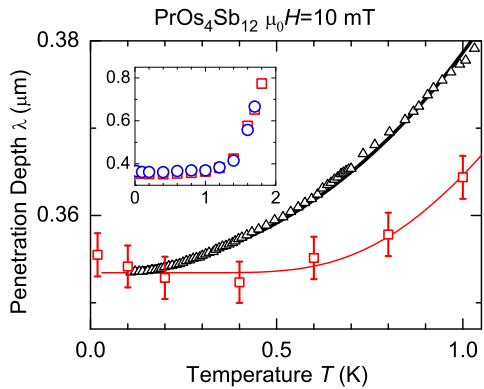}
 \caption{(color online) Temperature dependence of penetration depth~$\lambda$ in PrOs$_{4}$Sb$_{12}$ below $\sim$1~K\@. Squares: $\lambda_{\mu\mathrm{SR}}$ from GL model fitting. Curve: fit to BCS low-temperature expression~$\lambda(T) = \lambda(0)[1 + \sqrt{\pi\Delta/2T} e^{-\Delta/T}]$, $\lambda(0) = 0.3534(24)\ \mu\mathrm{m}$, $2\Delta/k_{B}T_{c}=4.9(1)$. Triangles: $\lambda_{\mathrm{rf}}$ from rf inductive measurements\cite{CSSS03} in the Meissner state (see text). Curve: fit to power law~$A + BT^n$, $n = 2.05(3)$. Inset: Temperature dependence of $\lambda_{\mu\mathrm{SR}}$ up to $T_{c}$.\cite{MSHS08} Squares: $\lambda_{\mu\mathrm{SR}}$ as in main panel. Circles: $\lambda_{\mu\mathrm{SR}}$ from Gaussian fits and Eq.~(\ref{eq:lambda}).}
 \label{fig:lambda}
 \end{center}
\end{figure}

Figure~\ref{fig:lambda} compares $\lambda_{\mu\mathrm{SR}}(T)$ in 
PrOs$_{4}$Sb$_{12}$ at low temperatures, obtained from GL model fits to TF-$\mu$SR data taken at $\mu_0H = 10$~mT, with the penetration depth increase~$\Delta\lambda_{\mathrm{rf}}(T) = \lambda(T) - \lambda(0)$ in the Meissner state obtained from rf inductive measurements.\cite{CSSS03} The rf data do not determine $\lambda(0)$ separately; in Fig.~\ref{fig:lambda} the quantity plotted (triangles) is $\lambda_{\mathrm{rf}}(T) = \Delta\lambda_{\mathrm{rf}}(T) + \lambda_{\mu\mathrm{SR}}(0)$, which allows direct comparison of the temperature dependencies from the two experiments. 

It can be seen that $\lambda_{\mu\mathrm{SR}}(T)$ is nearly constant below $\sim$0.6~K, indicative of a fully gapped quasiparticle excitation spectrum. As previously noted\cite{MSHB02} the BCS low-temperature expression $\lambda(T)=\lambda(0)(1 + \sqrt{\pi\Delta/2T}\, e^{-\Delta/T})$ (curve in Fig.~\ref{fig:lambda}) gives a good fit to the TF-$\mu$SR data for $T\leq 0.5T_{c}$, suggesting that the energy gap is isotropic. The increase of $\lambda_{\mathrm{rf}}(T)$ (triangles) with increasing temperature is much stronger than the increase of $\lambda_{\mu\mathrm{SR}}(T)$ and follows a power law~$A + BT^n$, $n \approx 2$ (Ref.~\onlinecite{CSSS03}). This is the discrepancy between the measurements noted above; a possible resolution is discussed in Sec.~\ref{sect:mbsc}.

As a test of the sensitivity of $\lambda_{\mu\mathrm{SR}}$ to the data fitting procedure, the functional form $G(t)\cos(\omega t+\phi)$, with a damped Gaussian envelope
\begin{equation}
 \label{eq:dG}
 G(t)=e^{-Wt}\exp\left(-\textstyle\frac{1}{2}\sigma^{2}t^{2}\right),
\end{equation}
was also used to fit the TF-$\mu$SR data from both the normal and superconducting states. In the normal state, where this function provides a good fit, the Gaussian relaxation arises from quasistatic Sb nuclear dipolar fields. The exponential damping increases with increasing field and is mainly due to normal-state susceptibility inhomogeneity; dynamic fluctuations of hyperfine-enhanced $^{141}$Pr nuclear spins\cite{SMAT07} play a small role at low fields. Below $T_{c}$ the exponential rate $W$ was fixed at the normal-state value for each field, so that the temperature dependence of the Gaussian rate $\sigma$ reflects the effect of the superconducting state. Some such procedure is necessitated by the strong statistical correlation between $W$ and $\sigma$ in Eq.~(\ref{eq:dG}); the time constant and the shape of the relaxation function are influenced by both of these parameters, so that correlations between them can result from small systematic errors. The principal justification for this \textit{ad hoc} fixing of $W$ is the insensitivity of the superconducting-state results to details of the fitting function [Eq.~(\ref{eq:dG})] discussed below. 

Determination of the vortex-state field distribution width requires correction for the normal-state relaxation. We take the superconducting-state Gaussian rate $\sigma_{s}$ to be given by $\sigma_{s}^{2}=\sigma^{2}-\sigma_{n}^{2}$, where $\sigma_{n}$ is the normal-state rate.\cite{SBK00} 

Equation~(\ref{eq:distri}) relates the second moment $\delta B^{2}$ of field distribution to $\lambda$ in the London limit. The second moment of the corresponding $\mu^{+}$ frequency distribution is $\delta\omega^{2}=\gamma_{\mu}^{2}\delta B^{2}$, where $\gamma_{\mu}$ is the $\mu^{+}$ gyromagnetic ratio. Then the estimated penetration depth $\lambda$ from Eq.~(\ref{eq:distri}) is
\begin{equation}
 \label{eq:lambda}
 \lambda\ (\mu \mathrm{m})=0.328/\sqrt{\delta\omega\ (\mu \mathrm{s}^{-1})}.
\end{equation}
As noted above the rms width $\sigma_{s}$ of the best-fit Gaussian is not necessarily $\delta\omega$, so that replacement of $\delta\omega$ in Eq.~(\ref{eq:lambda}) by $\sigma_{s}$ is not completely justified. Nevertheless $\sigma_{s}$ should scale with $\delta\omega$, and within its range of validity Eq.~(\ref{eq:lambda}) should give the correct temperature dependence of $\lambda$. This is because under these circumstances effects of nonzero $\xi$ are restricted to the high-field tail of field distribution, which is not heavily weighted in a Gaussian fit (cf. Fig. 1 of Ref.~\onlinecite{Kado04}). PrOs$_{4}$Sb$_{12}$ is a strongly type-II superconductor (GL $\kappa=\lambda/\xi \approx 30$, Refs.~\onlinecite{BFHZ02} and \onlinecite{MSHB02}), and this picture should be applicable. The temperature dependencies of $\lambda_{\mu\mathrm{SR}}$ obtained from Gaussian fits and Eq.~(\ref{eq:lambda}) (circles) and from GL model fits (squares)\cite{SMHC06} are shown in the inset of Fig.~\ref{fig:lambda} for an applied field of 10~mT\@. The results agree very well, giving strong evidence that $\lambda_{\mu\mathrm{SR}}$ is robust with respect to very different fitting strategies. 

\begin{figure}[ht]
 \begin{center}
 \includegraphics[width=0.45\textwidth]{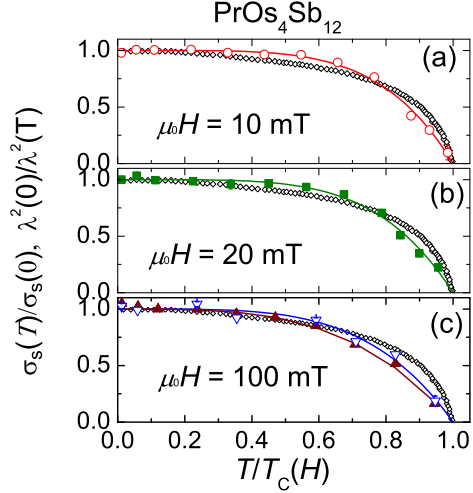}
 \caption{(color online) Normalized corrected superconducting-state $\mu^+$ spin relaxation rates~$\sigma_s(T)/\sigma_s(0)$ in PrOs$_{4}$Sb$_{12}$. (a): applied field $\mu_{0}H = 10$~mT (circles). (b): $\mu_{0}H = 20$~mT (squares). (c): $\mu_{0}H = 100$~mT (open triangles). Solid triangles: 100~mT data from Ref.~\onlinecite{HSKO07}. Curves: fits of the power law~$\sigma(T)/\sigma(0) = \left[1-(T/T_{c})^{n}\right] $ to the TF-$\mu$SR data. Diamonds: inverse-square Meissner-state penetration depth $\lambda^{2}(0)/\lambda^{2}(T)$ from rf data (Ref.~\protect\onlinecite{CSSS03}).}
 \label{fig:rate}
 \end{center}
\end{figure}

Figure~\ref{fig:rate} gives the temperature dependence of the corrected 
superconducting-state $\mu^{+}$ spin relaxation rate~$\sigma_{s}(T)$, normalized to the zero-temperature value~$\sigma(0)$, for $\mu_{0}H = 10$, 20, and 100~mT\@. The relaxation rates are well described by a power-law temperature dependence~$\sigma(T) = \sigma(0)[1-(T/T_{c})^{n}]$ (curves in Fig.~\ref{fig:rate}) over the entire temperature range $T<T_{c}(H)$, with the parameter values given in Table~\ref{table:power-law}\@.  
\begin{table}[ht]
\caption{Parameters from power-law fits $\sigma_s(T) = \sigma_s(0)[1-(T/T_{c})^{n}]$ to the corrected superconducting-state $\mu^+$ spin relaxation rate~$\sigma_s(T)$ in PrOs$_{4}$Sb$_{12}$.}\label{table:power-law}
\begin{ruledtabular}
\begin{tabular}{@{\quad}cccc}
 Field (mT) & $\sigma_{s}(0)$ ($\mu$s$^{-1}$) & $T_{c}(H)$ (K) & Exponent $n$\\
 \hline
 10 & 0.82(1) & 1.83(1) & 4.5(3)\\
 20 & 0.77(1) & 1.79(2) & 4.3(4)\\
 100 & 0.80(2) & 1.68(4) & 4.1(6)\\
 100\footnote{From Ref.~\protect\onlinecite{HSKO07}.} & 0.75(2) & 1.80(2) & 2.9(3)\\
\end{tabular}
\end{ruledtabular}
\end{table}
For all of the present data $n \gtrsim 4$, and for $\mu_{0}H = 10$ and 20~mT $\sigma_{s}(T)$ is nearly temperature-independent below $\sim$0.8~K\@. These features suggest the absence of gap nodes~\cite{SBK00,Amat97,MSHB02} at these fields. We should note, however, that such power-law fits have no clear physical interpretation and, furthermore, are dominated by the behavior of $\sigma_{s}(T)$ at higher temperatures, which is not the concern of this article. The large values of $n$ are merely signatures of the near temperature independence at low temperatures.

The normalized inverse-square penetration depth $\lambda^{2}(0)/\lambda^{2}(T)$ from rf measurements using $\lambda(0) = 0.353~\mu$m (Fig.~\ref{fig:lambda}), which is equal to $\sigma_{s}(T)/\sigma_{s}(0)$ in London superconductor as noted above, is also shown in Fig.~\ref{fig:rate} for comparison. For low fields a small but clear discrepancy at low temperatures can be seen: the relaxation rates decrease significantly less rapidly than $\lambda^{2}(0)/\lambda^{2}(T)$ with increasing temperature. This is of course the same discrepancy shown in Fig.~\ref{fig:lambda}. As noted above the effective penetration depth from TF-$\mu$SR can be modified by vortex interactions at higher fields; \cite{Soni07} the decrease of $n$ with increasing field (Table~\ref{table:power-law}) may reflect such an effect. 

Figure~\ref{fig:rate}(c) compares our TF-$\mu$SR data for $\mu_{0}H = 100$~mT with those from Ref.~\onlinecite{HSKO07} taken at the same field. There is reasonable agreement between all the data at low temperatures, but the exponential damping [Eq.~(\ref{eq:dG})] increases with field and begins to dominate the relaxation, thereby increasing the error in the Gaussian relaxation rate. The resultant scatter in the $\mu$SR data leads to noticeable differences in fit-parameter values between the two experiments (Table~\ref{table:power-law}), so that the situation for $\mu_{0}H = 100$~mT is uncertain. 

These results, together with a previous TF-$\mu$SR study on a different sample,\cite{MSHB02} indicate a generally flatter temperature dependence of $\lambda_{\mu\mathrm{SR}}(T)$ at low temperatures compared to that of $\lambda_{\mathrm{rf}}(T)$. As a whole the data suggest that the $\lambda_{\mu\mathrm{SR}}$-$\lambda_{\mathrm{rf}}$ discrepancy is intrinsic and robust for low fields ($\lesssim 20$~mT), where the $\mu$SR penetration depth determination is most reliable.\cite{Soni07} 

It can also be seen in Table~\ref{table:power-law} that the zero-temperature relaxation rate is essentially independent of field. In an isotropic superconductor such as cubic PrOs$_{4}$Sb$_{12}$ vortex-lattice disorder is expected to increase the low-field rate; increasing field (increasing vortex density) then decreases the rate as intervortex interactions stabilize the lattice.\cite{SBK00,NBHK02} Thus the field independence of the low-temperature rate indicates a substantially ordered vortex lattice at all fields, in which case the temperature dependence of the rate is controlled solely by the temperature dependence of the effective penetration depth. 

\subsection{\label{sect:sus}Magnetic susceptibility}

\begin{figure}[ht]
 \begin{center}
 \includegraphics[width=0.45\textwidth]{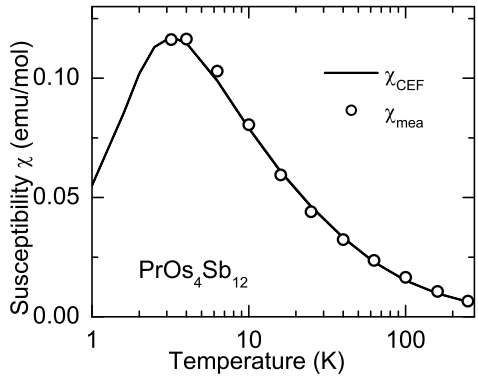}
 \caption{Temperature dependence of measured bulk susceptibility (circles)
 and best fit to model of interacting CEF-split Pr$^{3+}$ ions
 (curve) in PrOs$_{4}$Sb$_{12}$.
 }
 \label{fig:chi}
 \end{center}
\end{figure}

The temperature dependence of the normal-state magnetic susceptibility of the PrOs$_{4}$Sb$_{12}$ sample used in the TF-$\mu$SR experiments has been determined using a commercial SQUID magnetometer. Figure~\ref{fig:chi} shows the measured 
bulk susceptibility (circles). The data were fit to the model of CEF-split Pr$^{3+}$ ions with cubic $O_{h}$ symmetry interacting via a molecular field as discussed in Sec.~\ref{sect:CEF}. The resulting temperature dependence is given by the curve in Fig.~\ref{fig:chi}. The fit values of the CEF parameters in Eq.~(\ref{eq:hami1}) are listed in Table~\ref{table-B}. 
\begin{table*}[ht]
\caption{Fit values of CEF parameters in meV for PrOs$_{4}$Sb$_{12}$. $O_{h}$: cubic symmetry (no $\mu^{+}$ perturbation). $C_{4v}$: tetragonal symmetry ($\mu^{+}$ perturbation, cf. Sec.~\ref{sect:CEF}).}\label{table-B}  
\begin{ruledtabular}
\begin{tabular}{lccccccccc}
 \multicolumn{2}{c}{$B_{2}^{0}$} & \multicolumn{2}{c}{$B_{4}^{0}$} & \multicolumn{2}{c}{$B_{4}^{4}$ } & \multicolumn{2}{c}{$B_{6}^{0}$ } & \multicolumn{2}{c}{$B_{6}^{4}$} \\
 $O_{h}$ & $C_{4v}$ & $O_{h}$ & $C_{4v}$ & $O_{h}$ & $C_{4v}$ & $O_{h}$ & $C_{4v}$ & $O_{h}$ & $C_{4v}$ \\
\hline
 0 & -0.18 & 0.0154 & 0.0153 & 0.0771 & 0.0771 & 0.0007 & 0.0007 & -0.0154 & -0.0154\\
\end{tabular}
\end{ruledtabular}
\end{table*}
The fit value of the molecular field parameter $\ell$ was found to be $-1.67$~mole/emu, which is close to the value of $-2.54$~mole/emu found by Tayama \textit{et al}.\cite{TSSA03}

\subsection{\label{sect:knight}Normal-state muon Knight shift} 

\begin{figure}[ht]
 \begin{center}
 \includegraphics[width=0.45\textwidth]{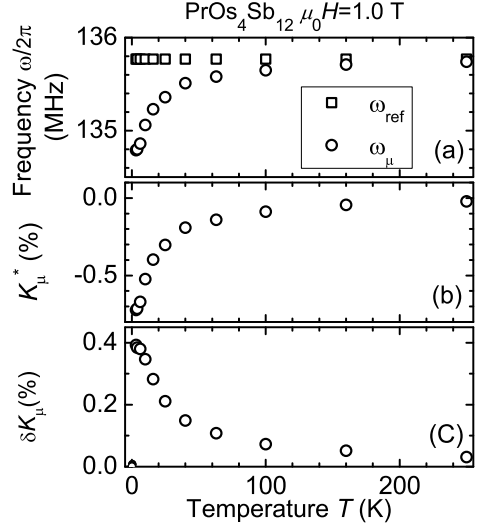}
 \caption{(a): Temperature dependencies of $\mu^{+}$ precession frequencies in PrOs$_{4}$Sb$_{12}$ ($\omega_{\mu}$) and an silver (Ag) reference ($\omega_{\mathrm{ref}}$). (b): Temperature dependence of $\mu^{+}$ frequency shift $K_{\mu}^{*}$ in the normal state. (c): Temperature dependence of relative rms linewidth $\delta K_{\mu}=\sigma/\omega_{\mathrm{ref}}$, where $\sigma$ is the $\mu^{+}$ Gaussian relaxation rate.}
 \label{fig:pos_fre}
 \end{center}
\end{figure}

We have performed TF-$\mu$SR experiments at applied field $\mu_{0}H=1.0$ T in the normal state of PrOs$_{4}$Sb$_{12}$ . Since the $\mu^{+}$ frequency shift is proportional to the magnetic field at the $\mu^{+}$ site, it is resolved better at higher fields. Fig.~\ref{fig:pos_fre}(a) shows the temperature dependence of the muon-spin precession frequencies from the sample and an Ag reference. Silver was used because Ag nuclear dipole fields are weak, leading to a well-defined reference frequency (narrow line), and the $\mu^{+}$ Knight shift in Ag is known (94 ppm).\cite{ScGy95} One can extract the $\mu^{+}$ frequency to an accuracy of $\sim$100 ppm.\cite{FAGG95a} The temperature dependence of the relative $\mu^{+}$ frequency shift $K_{\mu}^{\ast}$ is shown in Fig.~\ref{fig:pos_fre}~(b).

\begin{figure}[ht]
\begin{center}
\includegraphics[width=0.45\textwidth]{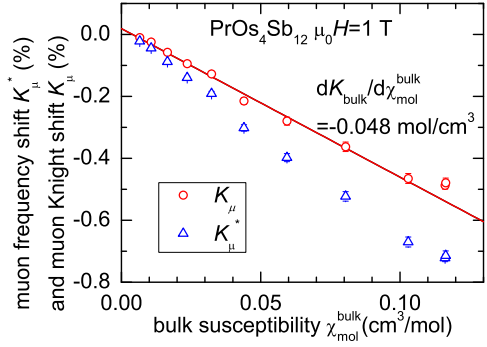}
\caption{(color online) Dependence of $\mu^{+}$ frequency shift $K_{\mu}^{*}$ (triangles) and corrected $\mu^{+}$ Knight shift $K_{\mu}$ (circles) on bulk molar susceptibility $\chi_{\mathrm{mol}}^{\mathrm{bulk}}$ in PrOs$_{4}$Sb$_{12}$, applied field $\mu_{0}H=1$~T\@. The straight line $K=K_{\mathrm{bulk}}$ (see text) is a fit to the data in the region $6.3\ \mathrm{K} \le T \le 250$~K\@.}
\label{fig:pos_knight}
\end{center}
\end{figure}

Figure~\ref{fig:pos_knight} (triangles) gives the dependence of $K_{\mu}^{\ast}$ on the bulk molar susceptibility $\chi_{\mathrm{mol}}^{\mathrm{bulk}}$ in PrOs$_{4}$Sb$_{12}$ in the normal state, with temperature an implicit variable. $K_{\mu}^{*}$ and $\chi_{\mathrm{mol}}^{\mathrm{bulk}}$ were measured in the same sample. As discussed in Sect.~\ref{sect:Knight-intro}, $K_{\mu}^{*}$ should be corrected for the effect of the Lorentz and demagnetization fields $4\pi(1/3 - D)M$, where $M$ is the magnetization. The demagnetization factor $D$ was estimated from (i) the ratio of the height and width of the samples,\cite{AkGa92} with the magnetic field applied perpendicular to the flat faces of the rectangular crystallites, and (ii) the ratio $f$ of individual crystal volume to the sample as a whole, taking spaces between the crystallites into account. We estimate the demagnetization factor for the entire sample $D_{\mathrm{samp}} = 0.824$, the demagnetization factor for individual crystals $D_{\mathrm{crys}} = 0.365$,\cite{AkGa92} and $f = 0.95$. Then the effective value is $D=fD_{\mathrm{samp}} + (1-D)n_{\mathrm{crys}}=0.80$.\cite{CBK77} Therefore, $K_{\mathrm{DL}} = 4\pi(1/3 - n)\chi_{V}^{\mathrm{bulk}}=-4\pi(0.47)\chi_{V}^{\mathrm{bulk}}$, where $\chi_{V}^{\mathrm{bulk}}$ is the bulk susceptibility per unit volume, or $K_{\mathrm{DL}}=-0.0243\chi_{\mathrm{mol}}^{\mathrm{bulk}}$. We subtract this from $K_{\mu}^{\ast}$ to obtain the corrected $\mu^{+}$ Knight shift $K_{\mu}$ (Fig.~\ref{fig:pos_knight}, circles).
A linear relation is obtained for $\chi_{\mathrm{mol}}^{\mathrm{bulk}}\lesssim 8.0\times 10^{-2}$ emu mole$^{-1}$ ($T\gtrsim 6.3$ K). This implies that at high temperatures the $\mu^{+}$ shift samples the same electrons that produce the large temperature-dependent bulk susceptibility component. Such behavior is generally expected from the ``bulk" Knight shift $K_{\mathrm{bulk}}$ in the absence of local perturbations. Only a small deviation from linearity (a single point) appears below $6.3$~K\@.

As discussed in Sec.~\ref{sect:Knight-intro}, the average shift arises solely from the hyperfine contact field at the interstitial $\mu^{+}$ for a cubic lattice. The dipole-dipole interaction, however, may split or broaden the $\mu^{+}$ line. Assuming the most probable $\mu^{+}$ stopping site ($\frac{1}{2}$, 0, 0.15) determined by Aoki \textit{et al}.,\cite{ATKS03}, we obtain a calculated dipolar Knight shift tensor with the following principal values for field directions ($x, y, z$) (cf.\ first column of Table~\ref{table-A}): $K_{\mathrm{dip}}^{x} = -2.70\times 10^{-2}\,\chi_{\mathrm{mol}}$, $K_{\mathrm{dip}}^{y} = 1.121\times 10^{-2}\,\chi_{\mathrm{mol}}$, and $K_{\mathrm{dip}}^{z} = 1.58\times 10^{-2}\,\chi_{\mathrm{mol}}$. Thus in the absence of additional broadening the $\mu^{+}$ spectrum should be split into three lines of equal weight. The observed relative rms linewidth $\delta K_{\mu}=4.68\times 10^{-2}\,\chi_{\mathrm{mol}}$ (data of Fig.~\ref{fig:pos_fre}(c); dependence on $\chi_{\mathrm{mol}}$ not shown) is, however, larger than the rms spread $\Delta K_{\mathrm{dip}}=\sqrt{[(K_{\mathrm{dip}}^{x})^{2}+(K_{\mathrm{dip}}^{y})^{2}+(K_{\mathrm{dip}}^{z})^{2}]/3}=1.92\times 10^{-2}\,\chi_{\mathrm{mol}}$. Thus the dipolar splitting cannot be resolved, although it contributes significantly to the linewidth.

\section{Discussion} \label{sect:discussion}

\subsection{Multiband Superconductivity}\label{sect:mbsc}

We first consider the discrepancy in measured magnetic penetration depth between the TF-$\mu$SR and rf experiments. Recently, extreme MBSC was found in PrOs$_{4}$Sb$_{12}$ from heat transport measurements by Seyfarth et al.\cite{SBMF05, SBMB06} Their thermal conductivity and other data are explained by small and large gaps $\Delta_{s}$, $\Delta_{l}$ on different sheets of the Fermi surface, together with different Fermi velocities $v_{Fs}$, $v_{Fl}$ and coherence lengths $\xi_{s,l} \approx \hbar v_{Fs,l}/\Delta_{s,l}$. A crossover field $H_{c2}^{s}$, which corresponds to the overlap of the vortex core electronic structure due to the small-gap band, is given by $H_{c2}^{s} = \Phi_{0}/2\pi\xi_{s}^{2} \approx 10$~mT,\cite{SBMB06} which is of the order of the lower critical field $H_{c1}$. Microscopic calculations of the local DOS in a two-band superconductor\cite{IMNM04} suggest that the small-gap band induces spatially extended QP states at low field. For $H > H_{c2}^{s}$, these loosely bound states overlap and become delocalized, with the dominant contribution to the DOS coming from the large gap band. In PrOs$_{4}$Sb$_{12}$ this high-field region includes most of the vortex state; it is in this sense that PrOs$_{4}$Sb$_{12}$ is an extreme multiband superconductor. 

Our TF-$\mu$SR measurements were performed for applied field $\mu_{0}H \ge 10~\mathrm{mT} \approx H_{c2}^{S}$. Then the small-gap states and their contributions to screening supercurrents are nearly uniform, the vortex-state field inhomogeneity is mainly due to large-gap suppercurrents, and $\lambda$ exhibits an activated temperature dependence if the large gap is nodeless. In contrast, the rf measurement of the surface penetration depth was performed in the Meissner state. Both large- and small-gap Cooper pairs contribute to the superfluid density. Its temperature dependence is controlled by both small- and large-gap superfluid densities; the small-gap contribution dominates the temperature dependence at low temperatures. Then the temperature dependence of the penetration depth in the vortex state from TF-$\mu$SR is weaker than that in the Meissner state from rf measurements. It should be noted that in this scenario the TF-$\mu$SR measurements give no information on the nodal structure of the small gap.

\subsection{\label{sect:LTKnight}Low-temperature Knight shift}
In Fig.~\ref{fig:pos_knight}, a small deviation from the linear relation of the $\mu^{+}$ Knight shift $K_{\mu}$ vs bulk paramagnetic susceptibility $\chi^{\mathrm{bulk}}$ appears below $6.3$~K in PrOs$_{4}$Sb$_{12}$. Kaplan, Schenck \textit{et al}.\ have reported deviations from linear $K$-$\chi$ relations from TF-$\mu$SR measurements of the $\mu^{+}$ Knight shift in single crystals of PrNi$_{5}$\cite{FAGG95a, FAGG95b,FAGS94} and PrIn$_{3}$.\cite{TAGG97} These were attributed to $\mu^{+}$-induced changes of the low-temperature susceptibility due to modification of the CEF of neighboring Pr$^{3+}$ ions. We argue that the $\mu^{+}$'s perturbing effect is small in PrOs$_{4}$Sb$_{12}$ for the following reasons:
\begin{itemize}

\item The deviation of $K_{\mu}(\chi)$ from linearity in PrOs$_{4}$Sb$_{12}$ is very small (Fig.~\ref{fig:pos_knight}). 

\item The superconducting transition temperature measured by TF-$\mu$SR is consistent with the bulk superconducting value (see Fig.~\ref{fig:rate}), so that this signature of superconductivity is not affected by the $\mu^{+}$ charge.

\item Pr$^{3+}$ ions are considerably more dilute in PrOs$_{4}$Sb$_{12}$ ($2.48 \times 10^{21}\ \mathrm{cm}^{-3}$) than in PrNi$_{5}$ ($11.8 \times 10^{21}\ \mathrm{cm}^{-3}$) or PrIn$_{3}$ ($9.81 \times 10^{21}\ \mathrm{cm}^{-3}$); the nearest-neighbor $\mu^{+}$-Pr$^{3+}$ distance is about 2~\AA\ longer in PrOs$_{4}$Sb$_{12}$ than in PrNi$_{5}$ or PrIn$_{3}$. Thus for comparable screening lengths the $\mu^{+}$ electric field at neighboring Pr$^{3+}$ sites in PrOs$_{4}$Sb$_{12}$ is more completely screened by the conduction electrons. Similarly, the experimental slope $dK_{\mathrm{bulk}}/d\chi_{\mathrm{mol}} = -4.8\times10^{-2}\ \mathrm{mol\ cm}^{-3}$ in PrOs$_{4}$Sb$_{12}$ (Fig.~\ref{fig:pos_knight}) is smaller than the corresponding values in PrNi$_{5}$ and PrIn$_{3}$ by more than an order of magnitude, indicating the RKKY coupling between $\mu^+$ spins and neighboring Pr$^{3+}$ ions is correspondingly weaker in PrOs$_{4}$Sb$_{12}$.

\item Schenck \textit{et al.}\cite{KGSB94} concluded from ZF-$\mu$SR studies that the spin fluctuations of the Pr$^{3+}$ electronic moments in PrNi$_{5}$ are slowed down and exhibit quasistatic behavior at low temperatures. They suggested that this behavior is related to $\mu^{+}$-induced modification of the low temperature susceptibility of the neighboring Pr ions. We did not observe such large widths of the magnetic field distribution at low temperatures [see Fig.~\ref{fig:pos_fre}(c)]. Our ZF- and LF-$\mu$SR measurements\cite{SMAT07} suggest instead that the observed dynamic $\mu^{+}$ relaxation is due to the $^{141}$Pr nuclear spin system with an hyperfine-enhanced effective nuclear moment.

\end{itemize}

We now discuss the small deviation from the linear $K$-$\chi$ relation that appears in Fig.~\ref{fig:pos_knight} below 6.3~K\@. It should be noted that muon-induced CEF modification is not the only mechanism for a nonlinear $K(\chi)$. It has been recognized for some time that such behavior, first observed in NMR Knight shifts,\cite{MacL81} might arise from effects such as (1)~temperature-dependent modification of the hyperfine coupling or (2)~the onset of a new susceptibility component at low temperatures. Thermal depopulation of excited CEF states at low temperatures, with consequent modification of the coupling,~\cite{MPL81} is an example of a Type-(1) mechanism. Nonlinear $K(\chi)$ behavior in heavy-fermion materials has recently been ascribed\cite{CYSP04} to a Type-(2) effect, related to a two-fluid description of the heavy-fermion state.\cite{NPF04} Our concern here is the possible influence of the muon charge on our vortex-state TF-$\mu$SR results in PrOs$_4$Sb$_{12}$; if other mechanisms were responsible for the small $K$-$\chi$ nonlinearity there would be no evidence for such influence. 

We therefore suppose the nonlinearity in PrOs$_4$Sb$_{12}$ is due solely to the muon's perturbing effect, i.e., the local modification of the Pr$^{3+}$ CEF splitting changes the local Pr$^{3+}$ susceptibility, so that the $\mu^{+}$ Knight shift is no longer monitoring the bulk susceptibility. Thus we can use the $\mu^+$ Knight shift data to determine this change. To obtain a quantitative determination, we use the approach of Schenck \textit{et al.},\cite{TAGG97,FAGG95a} making the following assumptions: (i) only the susceptibilities of the two nearest Pr$^{3+}$ ions are affected by the muon, and (ii) only the two nearest Pr$^{3+}$ ions contribute to the contact interaction at the $\mu^{+}$ site. Based on these assumptions, with the principal axes of $\chi$ chosen as the coordinate frame ($\mathbf{x},\mathbf{y},\mathbf{z}$), the Knight shift with the external field in the $i$ direction ($i=x, y$ or $z$) may be written as
\begin{equation}
 \label{eq:new Knight}
K^{i}=(A_{\mathrm{dip,NN}}^{ii}+A_{\mathrm{con}})\chi_{i}^{\mathrm{local}}+A_{\mathrm{dip,(1-NN)}}^{ii}\chi^{\mathrm{bulk}}.
\end{equation}
Here $\chi_{i}^{\mathrm{local}}$ is the altered susceptibility of the two nearest Pr$^{3+}$ ions, and $\chi^{\mathrm{bulk}}$ is the unperturbed bulk Pr susceptibility. The subscript~$\mathrm{NN}$ signifies that the sum in $A_{\mathrm{dip}}^{ii}$ only includes the two nearest Pr neighbors, and the subscript~$\mathrm{1-NN}$ indicates summation over all Pr$^{3+}$ ions in the Lorentz sphere other than the nearest neighbors. For the most probable $\mu^{+}$ site ($\frac{1}{2}$, 0, 0.15) (Ref.~\onlinecite{ATKS03}), the calculated values of $A_{\mathrm{dip}}$, $A_{\mathrm{dip,NN}}$, and $A_{\mathrm{dip},1-NN}$ are listed in Table~\ref{table-A}. 
\begin{table*}[ht]
\caption{Calculated values of $\mu^+$-Pr$^{3+}$ coupling constants $A_{\mathrm{dip}}$, $A_{\mathrm{dip},1-NN}$, and $A_{\mathrm{con}}$ (see text) in PrOs$_{4}$Sb$_{12}$.} \label{table-A}
\begin{ruledtabular}
\begin{tabular}{lcccc}
 $i$ & $A_{\mathrm{dip}}^{ii}$ (mol/cm$^{3}$) & $A_{\mathrm{dip,NN}}^{ii}$ (mol/cm$^{3}$) & $A_{\mathrm{dip},1-NN}^{ii}$ (mol/cm$^{3}$) & $A_{\mathrm{con}}$ (mol/cm$^{3}$) \\
 \hline
 $x$ & $-2.70\times 10^{-2}$ & $-5.08\times 10^{-2}$ & 2.37$\times 10^{-2}$ & \\
 $y$ & 1.121$\times 10^{-2}$ & 2.90$\times 10^{-2}$ & -1.78$\times 10^{-2}$ & 4.8$\times 10^{-2}$ \\
 $z$ & 1.58$\times 10^{-2}$ & 2.18$\times 10^{-2}$ & $-0.60\times 10^{-2}$ & \\
\end{tabular}
\end{ruledtabular}
\end{table*}
We take $A_{\mathrm{con}}$ to be the slope of $K_{\mu}$-$\chi$ in Fig.~\ref{fig:pos_knight}, since the tensor $\bm{\mathsf{A}}_{\mathrm{dip}}$ is traceless and the sum~$\sum_i A_{\mathrm{dip}}^{ii}$ vanishes as can be seen from Table~\ref{table-A}.

\begin{figure}[ht]
 \begin{center}
 \includegraphics[width=0.45\textwidth]{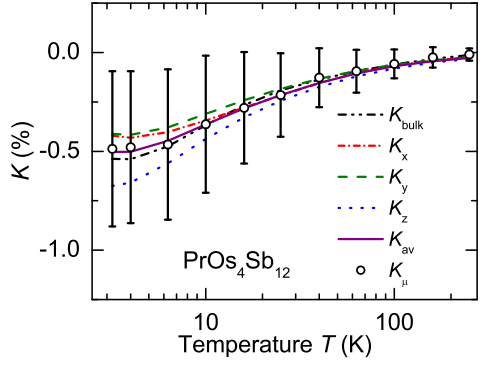}
 \caption{(color online) Temperature dependence of $\mu^{+}$ Knight shift in PrOs$_{4}$Sb$_{12}$. Circles: $\mu^{+}$ Knight shift $K_{\mu}$ from fig.~\ref{fig:pos_knight}. ``Error bars:" Gaussian linewidth $\delta K_{\mu}$ [Fig.~\ref{fig:pos_fre}(c)]. Dash-dot curve: calculated $K_{x}$. Dashed curve: calculated $K_{y}$. Dot curve: calculated $K_{z}$. Solid curve: average shift~$K_{\mathrm{av}}=(K_{x}+K_{y}+K_{z})/3$. Dash-dot-dot curve: $K_{\mathrm{bulk}}$ from linear fit (Fig.~\ref{fig:pos_knight}), expected for no $\mu^{+}$ perturbing effect.}
 \label{fig:pos_knight_cef}
 \end{center}
\end{figure}

\begin{figure}[pt]
 \begin{center}
 \includegraphics[width=0.45\textwidth]{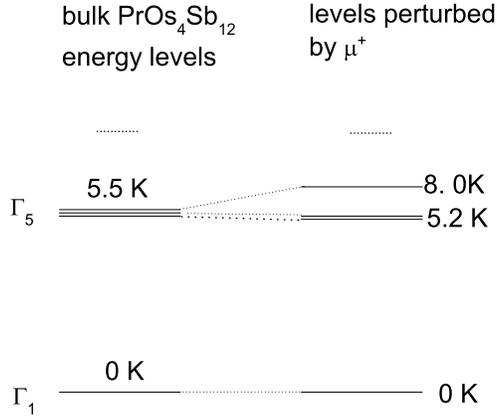}
 \caption{Original bulk energy-level scheme of the Pr ions in PrOs$_{4}$Sb$_{12}$, compared to the local level scheme induced by $\mu^{+}$.}
 \label{fig:energy}
 \end{center}
\end{figure}

In order to obtain the values of $\chi_{i}^{\mathrm{local}}$, we must modify ${\mathcal{H}}_{\mathrm{CEF}}$ to produce the required level changes. In the presence of a muon, the symmetry of the neighboring Pr$^{3+}$ ions is not the original cubic symmetry, but rather a very approximate tetragonal ($C_{4v}$) symmetry. The general Hamiltonian describing this tetragonal symmetry is\cite{FAGG95a,FAGG95b}
\begin{equation}\label{eq:hami}
{\mathcal{H}}_{\mathrm{CEF,tet}} = B_{2}^{0}O_{2}^{0}+B_{4}^{0}O_{4}^{0}+B_{4}^{4}O_{4}^{4}+B_{6}^{0}O_{6}^{0}+B_{6}^{4}O_{6}^{4},
\end{equation}
which has five free parameters, compared to two in the original cubic Hamiltonian. In principle, the muon may modify the charge distribution around the Pr ions in every direction. However, if we only consider changes occurring along the Pr$^{3+}$-$\mu^{+}$ axis for simplicity, it has been shown that $B_{4}^{4}$ and $B_{6}^{4}$ remain at their cubic values.\cite{TAGG97} This reduces the number of the free parameters from five to three. Hence, using Eq.~(\ref{eq:new Knight}), we can fit the $\mu^{+}$ Knight shift by choosing appropriate values of $B_{2}^{0}$, $B_{4}^{0}$, and $B_{6}^{0}$. Our theoretical fits in PrOs$_{4}$Sb$_{12}$ are shown in Fig.~\ref{fig:pos_knight_cef}. 
The anisotropy is consistent with the new CEF and Eq.~(\ref{eq:new Knight}), decreases with increasing temperature, and is well within the spread~$\delta K_{\mu}$ of $\mu^{+}$ shifts (shown as ``error bars'' in Fig.~\ref{fig:pos_knight_cef}). The data are very well described by the average shift~$K_{\mathrm{av}}=(K_{x}+K_{y}+K_{z})/3$, which is almost the same as $K_{\mathrm{bulk}}$ from Fig.~\ref{fig:pos_knight} that assumes no $\mu^{+}$ perturbing effect. The fitted CEF parameters are listed in Table~\ref{table-B} under the columns with heading $C_{4v}$. The changes in the parameters, which are very small and probably not statistically significant, give rise to a correspondingly small rearrangement of the local energy levels. A comparison of bulk and locally perturbed energy level scheme of PrOs$_{4}$Sb$_{12}$ is shown in Fig.~\ref{fig:energy}. We see that the change of splitting between the ground state and first excited state energies is only 0.3~K, and conclude that the perturbing effect of the $\mu^{+}$ charge in PrOs$_{4}$Sb$_{12}$ is negligibly small.

\section{Conclusions} \label{sect:conc}

In this paper we have presented TF-$\mu$SR results in the first Pr-based heavy-fermion superconductor PrOs$_{4}$Sb$_{12}$. It is found experimentally that the effective penetration depth $\lambda_{\mathrm{eff}}$ is temperature-independent in the vortex state for low temperatures, consistent with a nonzero gap for quasiparticle excitations. In contrast, the temperature dependence of the penetration depth from rf inductive measurements suggests point nodes in the gap. This discrepancy can be resolved in a scenario based on the recent discovery of two-band superconductivity in PrOs$_{4}$Sb$_{12}$.

The temperature dependence of the normal-state $\mu^{+}$ Knight shift in PrOs$_4$Sb$_{12}$ reveals a linear scaling of the Knight shift with the bulk magnetic susceptibility at high temperatures. A very small deviation from the linear relation appears in PrOs$_4$Sb$_{12}$ below $6.3$~K\@. Such deviations have been explained by $\mu^{+}$ induced modifications of the susceptibility of neighboring Pr$^{3+}$ ions due to a change of the CEF splitting. Our data indicate, however, that this modification is very small in PrOs$_4$Sb$_{12}$. A model calculation based on CEF theory and the associated perturbed electronic energy levels confirms the smallness of the $\mu^{+}$ perturbation effect. Therefore it is unlikely that the discrepancy between TF-$\mu$SR and rf inductive measurements is caused by the $\mu^{+}$ charge or, more generally, that the $\mu^{+}$ charge influences the role of the Pr$^{3+}$ CEF states in the superconductivity of PrOs$_4$Sb$_{12}$.

\begin{acknowledgments}
We are grateful for technical assistance from the TRIUMF Centre for Molecular and Materials Science during the experiments. This work was supported in part by the U.S. NSF under grants 0422674 (Riverside), 0604015 (Los Angeles) and 0335173 (San Diego), by the Canadian NSERC and CIAR (Burnaby), and by the U.S. DOE under grant DE-FG-02-04ER46105 (San Diego). Work at Los Alamos was performed under the auspices of the U.S. DOE.
\end{acknowledgments}




\end{document}